# Hierarchical modularity of nested bow-ties in metabolic networks

Jing Zhao[1,2,4], Hong Yu[2], Jian-Hua Luo[1], Z. W. Cao[2§], Yi-Xue Li [2,3,1§]

[1]School of Life Sciences & Technology, Shanghai Jiao Tong University, Shanghai 200240, China
[2]Shanghai Center for Bioinformation and Technology, Shanghai 200235, China
[3]Shanghai Institutes for Biological Sciences, Chinese Academy of Sciences, Shanghai 200031, China
[4]Department of Mathematics, Logistical Engineering University, Chongqing 400016, China

[§]Corresponding author

Email addresses:
    JZ: zjane_cn@sjtu.edu.cn
    HY: yuhong@scbit.org
    JHL: jhluo@sjtu.edu.cn
    ZWC: zwcao@scbit.org
    YXL: yxli@sibs.ac.cn

# Abstract

**Background**

The exploration of the structural topology and the organizing principles of genome-based large-scale metabolic networks is essential for studying possible relations between structure and functionality of metabolic networks. Topological analysis of graph models has often been applied to study the structural characteristics of complex metabolic networks.

**Results**

In this work, metabolic networks of 75 organisms were investigated from a topological point of view. Network decomposition of three microbes (*Escherichia coli*, *Aeropyrum pernix* and *Saccharomyces cerevisiae*) shows that almost all of the sub-networks exhibit a highly modularized bow-tie topological pattern similar to that of the global metabolic networks. Moreover, these small bow-ties are hierarchically nested into larger ones and collectively integrated into a large metabolic network, and important features of this modularity are not observed in the random shuffled network. In addition, such a bow-tie pattern appears to be present in certain chemically isolated functional modules and spatially separated modules including carbohydrate metabolism, cytosol and mitochondrion respectively.

**Conclusions**

The highly modularized bow-tie pattern is present at different levels and scales, and in different chemical and spatial modules of metabolic networks, which is likely the result of the evolutionary process rather than a random accident. Identification and analysis of such a pattern is helpful for understanding the design principles and facilitate the modelling of metabolic networks.

# Background

Cellular metabolism is an essential process for the maintenance of life and metabolic networks have been extensively studied [1-7]. Although a large variety of metabolic reactions can be found in different organisms, metabolic networks are highly conserved across them. It remains a highly interesting and challenging problem to understand the architectural characteristics and "design" principles of the metabolic networks in relation to their function. An important finding is that metabolic networks, as well as other real-world complex networks, have topologies that differ markedly to those found in simple randomly connected networks [8], which suggests that their non-random structures could imply significant organizing principles of metabolic networks.

Cellular functions are carried out in a modular way [9], while it has been suggested that biological modules correlates with locally dense links in molecular networks [10]. Efforts have been directed towards the recognition of modules in metabolic networks using mathematical methods, especially graph theory. Graph-theoretic methods analyze networks from topological point of view using minimal prior knowledge about biological function, thus have the potential to give new insight into metabolism based on the unbiased structural modules. Newman and Girvan regarded a network has modularity (or community) structure if its nodes could be properly divided into groups within which the nodes are highly connected, but between which they are much less connected [11]. Several algorithms have been developed to break up a metabolic network into modules [12-16], while the corresponding modularity metric could be used as a quantitative criterion to evaluate the decomposing quality of a network [11]. Guimera et al. further proposed that the modularity of networks must always be compared to the null case of a random graph. They regarded a network has statistically significant modularity if its modularity metric, i.e., the largest modularity metric of all possible partitions of the network, is statistically bigger than that of the randomised counterparts [17]. They also developed a simulated annealing algorithm to compute the largest modularity metric of a network [14, 18]. On the other hand, it has been discovered, that the global metabolic network is organized in the form of a bow-tie [19]. On the basis of such a bow-tie topology, Ma and Zeng have proposed a decomposing algorithm based on the shortest path combined with "majority rule" [16]. These methodologies are useful for the analysis of the design principle of metabolic network.

In this work, 75 metabolic networks were constructed from organisms including 8 eukaryote, 56 bacteria and 11 archaea. By applying the decomposing method similar to [16], the topological features of various graph models were studied at different levels, sub-cellular localizations, and biochemical pathways in the form of bow-tie. To mine the inherent topology of metabolic networks, the features from *E.coli* network were then compared with those of the properly randomized counterparts that preserve the linkage degree of each node and the total number of directed and bi-directed arcs [20, 21].

## Results/Discussion

### Decomposing the metabolic networks

The metabolic networks of three microbes (*Escherichia coli*, *Aeropyrum pernix* and *Saccharomyces cerevisiae iND750*) were decomposed. Only the results of the *E.coli* network are presented in this paper for illustration. The other two networks displays similar features as the *E.coli* network and the relevant analysis results are provided in Additional file 1 (part II and III).

The metabolic network of *E. coli* K-12 MG1655 consists of 934 nodes and 1437 arcs. The largest connected part of this network embraces 575 nodes and its topology exhibits a bow-tie architecture consisting of four parts: giant strong component (GSC), substrate subset (S), product subset (P) and isolated subset (IS). To further decrease the complexity, the GSC part is reduced to a Core through the method of [19]. See part I of Additional file 1 for a visualization of the bow-tie structure for *E.coli* network.

The hierarchical clustering tree for the Core of the GSC, obtained by our decomposition algorithm, is shown in Figure 1. According to the modularity metric from Newman and Girvan [11], 12 clusters of the Core appeared as shown in Figure 2, in which the nodes belonging to the same cluster have the highest degree of structural equivalent, i.e., the clusters are still strongly connected (cluster 2,3,5,6,7,8 and 12), or most of the nodes are strongly connected (cluster 1,4,9,10 and 11). Figure 3 illustrates the decomposition of the whole metabolic network.

Matching modules to particular metabolic functions may reveal the biological significance of this modularity [16, 18]. Following Guimerà and Amaral [18], we mapped the modules to KEGG pathway [22, 23]. A cartographic representation [18] of the metabolic network is shown in Figure 4, in which each node corresponds to a cluster. The colours in Figure 4 represent different categories of metabolism while the coloured areas indicate the percentage of respective metabolism within the module.

Figure 4 illustrates that some modules generated by our algorithm are dominated by one major category of metabolisms. For instance, the reactions in the 3$^{rd}$ module are mainly carbohydrate metabolisms that include the majority of metabolites from TCA cycle with glyoxylate bypass, as shown in Figure 5. However, the majority are mixtures of pieces of several conventional biochemical pathways. Extreme examples are module 1 and module 5. When examining the nodes in them we found that there are heavily overlapping compounds both by carbohydrate and amino acid metabolism. It is difficult to assign these metabolites to a single module because they are playing dual and even multiple roles in several metabolism processes. For example, pyruvate in the 1$^{st}$ module is a key metabolite that connects the metabolism of carbohydrates, amino acids and the energy metabolism. See Figure S5 in Additional file 1 for different pathways grouped in module 5. On the other hand, a standard textbook pathway can break into several modules, which is especially true for the three central pathways – Embden-Meyerhof-Parnas (EMP), tricarboxylic acid (TCA) and pentose phosphate pathway (PPP). One possible explanation is that the metabolites in these central pathways are used as common precursors for biosynthesis of universal building blocks [24] and are thus placed in different modules. Figure 6 shows how the 12 common precursors scatter in different modules. This finding agrees with earlier observations concerning the high diversity of the TCA and EMP pathway [25, 26], as well as the clustering results for *E.coli* metabolic network obtained by other algorithms [13, 16, 27].

It can be seen that the modules generated from topology do not completely overlap with traditional biological pathways. However, this purely graph-theoretic clustering without any use of biology-specific details may provide new insight into metabolism and useful hints to the functional significance of those unknown reactions. Taking module 3 as an example, parts of amino acid metabolites and nucleotide biosynthesis metabolites are connected closely around TCA cycle in module 3, which provides convenient plugs interrelating with other modules. For instance, acting as an important interface to other modules, aspartate (ASP-L) provides amino bases for the synthesis of other categories of amino acid in module 4 and module 5, and is also used for pyrimidine synthesis in module 5; while 2-oxoglutarate (AKG), as the precursor of glutamate family, is clustered into module 9 as the plug between module 3 and module 9. Hopefully, more research will clarify the biological significance of the underlying difference between topological modules and traditional pathways.

### Hierarchical modularity of metabolic networks in the case of nested bow-tie

To further investigate the macroscopic structure of each sub-network, the node distributions in the bow-tie structure of the sub-networks were listed in Table 1. It can be seen that almost all of the twelve sub-networks have formed bow-tie structures, similar to the global network. We show the four parts of each sub-bowtie with distinct colours in Figure 3, while in Figure 7 we illustrate the corresponding sub-tree of module 3 and its bow-tie structure. The only exception is the 6$^{th}$ module, which doesn't have the P part, but closely related with module 5 and module 7. This could be caused by our decomposing algorithm that starts from decomposing the core of GSC, which contains highly abundant reversible reactions.

As pointed out in the method part, with the dissimilarity index defined in our algorithm, nodes that belong to the same sub-tree are not necessarily connected to each other, so are modules that correspond to the neighboured sub-trees. For example, although the sub-tree of module 8 is near that of module 7 in the hierarchical clustering tree shown in Figure 1, module 8 is not connected with module 7 but with module 4, 5 and 12, whose sub-trees are far away, as Figure 4 shows. Thus the hierarchical tree cannot reflect the actual linkage relation between modules. To mine how the small modules are organized into bigger ones, we drew a coarse-grained graph to illustrate the connections between the GSC parts of the sub-networks as Figure 8 shows. Each node in Figure 8 corresponds to a cluster, while two nodes in Figure 8 are defined as being connected if and only if the constituent nodes in corresponding GSC parts are linked. Such connecting topology is different from that in Figure 4, in which the arcs correspond to the links between the sub-networks.

It is thus noted from the definition of strongly connected graphs that, if some nodes in Figure 8 can be combined into a strongly connected sub-graph, the merger of the corresponding sub-networks may form a bigger bow-tie whose GSC is the union of the individual GSC parts [28]. For example, the unions of clusters, such as {1,2,3}, {1,3,4}, {5,8,10,11,12}, and the union of all the twelve clusters have bow-tie structures, but the following clusters {1,2}, {4,9,10}, {10,11,12} can't form strongly connected sub-graphs, thus are not bow-tie. In this way, different sub-networks of bow-tie structures can be combined to form bigger bow-ties at higher level. In other words, Figure 8 delineates the "roadmap" how little bow-ties are nested into larger bow-ties through their GSC parts.

The combination of different bow-ties was also compared with the global bow-tie from proportional scale. One hundred and fifty bow-ties were hierarchically generated by random combinations of a number of basic bow-ties from the coarse-grained graph in Figure 8. Their node distributions of the four parts (GSC, S, P, and IS) were listed in Table S3 of Additional file 1. The percentage discrepancies of the four parts to those of the global one were also summarized in Table S3. Interestingly, the node distribution of the nested bow-ties is approximately consistent with that of the global network with an average absolute error of 0.0854, which means each smaller bow-tie can be considered as a miniature of the global one.

In this sense, metabolic networks seem to be designed in such a way that many similar small modularized bow-tie units, which are hierarchically nested and reoccur at different scales and levels, are coupled level-by-level into a larger network.

### Comparison between the *E.coli* network and an ensemble of randomly connected networks

Comparing metabolic network with randomized counterparts could reveal intrinsic difference between them [17, 20, 21]. Sixty random networks were constructed by reshuffling the links of the *E.coli* metabolic network [20, 21]. The graph metrics of the 60 random networks are listed in Table S4 of Additional file 1 and the comparison with the *E.coli* network is summarized in Table 2. Topological analysis revealed that the macroscopic structures of the random networks still preserve a global bow-tie, but substantial difference exists between the *E.coli* metabolic network and randomized ones in term of the sizes of bow-tie parts, average clustering coefficient and modularity metric. It can be seen that the clustering coefficients of the random networks are almost equal to zero, in big contrast to that of the *E.coli* network. This clear difference implies an overall loose connection of randomized networks but a cliquish topology of the *E.coli* metabolic network [29]. Such different topological patterns are observably presented in their Cores, as Figure 9 shows. The Core of *E.coli* network exhibits distinct cohesive areas being sparsely linked together, while the randomized one is linked in such an approximately equal density that almost no clear-cut "cliques" appear within it.

Table 2 also shows that the modularity metric of the *E.coli* network is some 22 standard deviations above that of the randomized network. Thus the higher modularity of *E.coli* network is unlikely to arise at random. According to the scheme of Maslov et al. and Guimera et al. [17, 21], metabolic network exhibits significantly higher modularity compared with those randomized counterparts statistically. On the other hand, since a network with modularity metric higher than 0.7 could be thought as modular network [11], both *E.coli* network and its randomized version, whose values of modularity are 0.85 and 0.76 respectively, could be judged as being modular according to the module definition proposed by Newman and Girvan, i.e., the network can be broken up in such a way that it has as many as within-module links and as few as possible between-module links. However, modules defined like this and detected by simulated annealing are not compatible with the bow-tie like modules (see Figure S16 in Additional file 1). While decomposing this random network in the same way as the *E.coli* network by our algorithm, we found that, no matter how the hierarchical clustering tree was cut at different level, several unconnected sub-networks consisted of many isolated nodes would be generated. Figure 9(B) shows one manner of the partitions that generates 12 clusters, in which the nodes belonging to the same cluster are not linked together, and the modularity metric corresponding to this decomposition is computed as only 0.0612. In contrast, each cluster of the Core for *E.coli* network is connected (Figure 9(A)), which could lead to bow-tie pattern decomposition. The corresponding modularity metric of Figure 9(A) is 0.7062. That is to say, structural equivalent nodes tend to connect with each other in the real metabolic network, but split away in random networks. This phenomenon could be the result of local interactions within metabolic pathways.

In summery, in the sense of modules defined by Newman and Girvan, both *E.coli* network and random ones are modular organized. However, *E.coli* network has significantly higher modularity, and also could be decomposed as bow-tie modules, while the random network does not exhibit modular organization in terms of bow-tie modules. These comparative results indicate that the highly modularized bow-tie unit is an intrinsic and significant feature of metabolic networks, rather than a random phenomenon.

In addition, we have also seen that the use of degree-preserving rewiring here provides some views into the statistical and graph-theoretic differences between real metabolic networks and random ones. We believe that additional work is required to understand what the other distinctions are, and what the biological significance is of these differences.

**Bow-tie units from a chemical and spatial viewpoint**

Assuming that the highly modularized bow-tie unit is common to metabolic networks, the bow-tie may be observed from various systems such as cellular organelles and metabolic pathways of fundamental bio-molecules. Reactions of the three basic metabolisms, carbohydrate metabolism, lipid metabolism, and amino acid metabolism, were retrieved from the database of [30] for 75 organisms (Eukaryote: 8; Bacteria: 56; Archaea: 11). Our analysis shows that all the reactions in carbohydrate metabolism have framed a bow-tie structure for all the examined organisms, as being provided by part VII of Additional file 1. But neither the lipid reactions nor the amino acid reactions can build a bow-tie. One possible explanation to this could be the different roles they play. Carbohydrate metabolism plays a fundamental role in nutrients and energy metabolism, which produces lots of flexible intermediated metabolites for the biosynthesis of lipids, amino acid and other materials.

Similar analysis was done to the reactions of the *Saccharomyces cerevisiae* iND750 [31] according to sub-cellular localisations. Three cellular compartments, cytosol, mitochondrion, and peroxisome, were studied which include relatively more metabolites. It was found that the sub-networks of cytosol reactions and mitochondrion reactions could exhibit the bow-tie patterns as shown in Table 3. It is known that mitochondrion is functionally relatively independent organelle, while the majority of metabolic reactions take place in cytosol. That the peroxisome reactions do not form a bow-tie could be caused by the scarcity of reaction information of peroxisome. However, with the development of genomics, proteomics and metabonomics, and the accumulation of sub-cellular information of more reactions, we speculate that it is possible to find bow-tie structures in more organelles.

In brief, bow-tie pattern is also present in elementary metabolism such as carbohydrates, and in cellular compartments of mitochondria and cytosol. These results seem to indicate that the modularity of bow-tie patterns is common to metabolic networks. At the same time, the complete bow-tie patterns in mitochondria and carbohydrates pathways could also imply some independent functional clues.

**Significance of modularity in the form of bow-tie structure**

In the long evolutionary process of metabolism networks and their components, the structure of modularity could contribute significantly to the function of metabolic networks. Here, the recurrence of bow-tie structures suggests that bow-tie modules may act as another kind of building block of the genome-based metabolic network during the evolutionary process, indicating that evolution might copy and reuse existing modules to give rise to ever higher forms of complexity when new function calls for it.

Another contribution would be network robustness. It is argued that the GSC part in the bow-tie of the metabolic network is robust against mutations because there are multiple routes between any pair of nodes within the GSC [19, 32]. While a modular metabolic network which is nested by many relatively independent and robust bow-tie units, will provide more advantages in generating coordinated response to various stimuli from environment and further increase the robustness of the whole metabolic system.

Moreover, selection of bow-tie as a structural building unit seems to be a concise and smart option for constructing metabolic networks. From the standard biochemical point of view, the metabolic system is organized as a bow-tie whose knot is made up of a small handful of activated carriers and 12 precursors, with a large "fan-in" of nutrients, and a large "fan-out" of products in biosynthetic pathways[24, 33]. Such organization pattern has been reported to be present in various biological systems, such as in signal transduction systems, transcription and translation processes, and immune systems [24, 32-36]. The bow-tie model here could give alternative view of the biological metabolites flow from the topological aspects, where the knot is much thicker than that of above. We will refer to thin bow-tie and thick bow-tie to distinguish these two models. These two bow-tie models are similar in that they both specially identify and isolate the carriers. It is noted that, besides the carriers, the thin bow-tie model includes only the 12 precursors as its knot, whereas the thick bow-tie model here also contains these 12 precursors, but together with the three essential pathways – TCA (tricarboxylic acid) cycle, pentose phosphate pathways and glycolysis pathways, which generate the precursors, as well as much more metabolites and reactions. Although different bow-tie models in details, the similar organization pattern can both facilitate the kind of extreme heterogeneity that allows for robust regulation, manageable genome sizes and biochemically plausible enzymes [24].

The knot in our model denotes the most tightly connected part of the network and is comprised of concentrated intermediated metabolites. This thicker knot would possibly allow the network to manipulate flexible controls through the knot and provide more interfaces with inputs and outputs to meet an emergency or process new metabolites. On the other hand, the thicker knot may reveal the flexibility that the organism has in interchanging nutrients and products. *E. coli* in particular heavily uses products of other organism metabolism as nutrients, as do most organisms, but can also live on fairly minimal media as well. The thick knot may reflect this flexibility, but further research will be needed to full explain these connections.

## Conclusions

In this survey we have attempted to reveal the topological features of graph models from the view of the design principle of metabolic networks. Our results suggest that metabolic networks exhibit hierarchical modularity in the form of modularized bow-tie units, whereas this highly structured modularity is not present in random graphs with comparable statistical weight. This finding is consistent with the conclusions from a number of studies that these structures result from universal and fundamental organizing principles for efficiency and robustness, rather than frozen accidents of evolution. On the other hand, such nested bow-tie topology may also be the result of natural selection of biological evolution, which could be conceived as a process where the same patterns and processes repeat at each stage, and are nested at multilevel. The perspectives of this paper would provide useful hints for understanding the function and evolution of metabolic networks, as well as the modeling and simulation of complex biological systems.

# Methods

**Data preparation and network reconstruction**

In this study, the metabolic data were extracted from the database developed by Ma and Zeng based on the Kyoto Encyclopedia of Genes and Genomes (KEGG) [30]. In this database, the information concerning the reversible reactions was specified. In addition, some small molecules, such as adenosine triphosphate (ATP), adenosine diphosphate (ADP), nicotinamide adenine dinucleotide (NAD) and $H_2O$, are normally used as carriers for transferring electrons or certain functional groups and participate in many reactions, while typically not participating in product formation. Therefore, in order to reflect biologically relevant transformations of substrates, these kinds of small molecules, as well as their connections were manually excluded from the database when no products were formed from them. It should be noted that this method of exclusion is not determined by compounds, but by the reaction. For example, glutamate (GLU) and 2-oxoglutarate (AKG) are currency metabolites for transferring amino groups in many reactions, but in the following reaction:

AKG + $NH_3$ + NADPH = GLU + $NADP^+$ + $H_2O$,

AKG participates in producing GLU, i.e., they are primary metabolites. Hence the connections through them should be considered. A metabolic network reconstructed from this database is represented by a directed graph whose nodes correspond to metabolites and whose arcs correspond to reactions between these metabolites, in which irreversible reactions are presented as directed arcs while reversible ones as bi-directed arcs. For example, the irreversible reaction,

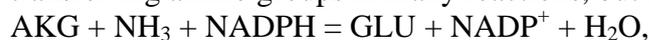
L-Glutamine + 2-Oxoglutarate $\rightarrow$ L-Glutamate,

corresponds to two directed arcs, i.e., L-Glutamine $\rightarrow$ L-Glutamate and
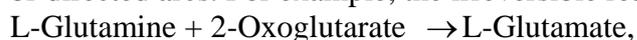
2-Oxoglutarate $\rightarrow$ L-Glutamate.

The metabolic network iND750, a fully compartmentalized genome-scale metabolic model of *Saccharomyces cerevisiae* constructed by Duarte et al. [31], was then studied. This set of data is the most complete metabolic data in the public domain that includes information on sub-cellular localization. Compartmentalization of information, which includes the localization to the cytosol, Golgi apparatus, mitochondrion, nucleus, endoplasmic reticulum, vacuole, peroxisome, or extracellular space, is given for each reaction. Reactions were assigned to the cytosol by default unless there was evidence that a metabolite was found in a particular compartment.

Referring all the reactions in iND750 to the database of [30], we manually removed the connections through currency metabolites, such as $H_2O$, ATP, NADH, thus the reconstructed network of *S. cerevisiae* is represented as a directed graph.

**Topological features and metrics of networks**

**Bow-tie structure**: A network with bow-tie structure consists of four parts: giant strong component (GSC), substrate subset (S), product subset (P) and isolated subset (IS) [19]. The GSC is the biggest of all strongly connected components and is much larger than all the other ones, while a strongly connected component is defined as the largest cluster of nodes within which any pair of nodes is mutually reachable from each other. S consists of nodes that can reach the GSC but cannot be reached from it, while P consists of nodes that are accessible from the GSC, but do not link back to it.

The IS contains nodes that cannot reach the GSC, and cannot be reached from it. The GSC part may include many linear branches, which consist of several reversible reactions. Removing these linear branches will lead to the Core of the GSC, which is still strongly connected [19]. See part I of Additional file 1 for visualization of bow-tie structure of *E.coli* network.

**Clustering coefficient**: the clustering coefficient of node *v* measures the extent that its neighbours are also linked together, i.e., to form a clique[29]:

$$CC(v) = \frac{2|N(v)|}{d(v)(d(v)-1)},$$

where $|N(v)|$ denotes the number of links between neighbours of node *v*, *d(v)* is the degree of node *v*. The value of *CC (v)* is between 0 and 1. To some extent, the clustering coefficient of a network, i.e., the average of *CC (v)* over all *v*, could reflect the cliquishness of the network [29].

**Modularity metric**: For a given decomposition of a network, the modularity metric is defined as the fraction of arcs within clusters minus the expected fraction of edges if the arcs are wired with no structural bias [11]:

$$M = \sum_{i=1}^{r}[e_{ii} - (\sum_{j} e_{ij})^2]$$

where *r* is the number of clusters, $e_{ij}$ is the fraction of arcs that leads between vertices of cluster *i* and *j*. Guimera et al. defined the modularity metric of a network as the largest modularity metric of all possible partitions of the network [17], and they also developed a simulated annealing algorithm to compute the modularity metric of a network [14, 18]. The simulated annealing algorithm identifies modules by maximizing the network's modularity parameter so that there are as many as within-module links and as few as possible between-module links.

When being used for different ways of partition of the same network, the modularity metric can measure which partition is better. If used for different networks, the largest modularity metric can measure their modular extents. In this study, we applied this metric in both of the different ways. When decomposing the network by our algorithm, we detected the best cut of the hierarchical tree based on this metric. While comparing *E.coli* network with its randomized counterparts, we measured the modular extent of the network by its largest modularity metric computed by simulated annealing algorithm.

**Algorithm to decompose the genome-based metabolic network**

Usually, clustering of a graph starts with a dissimilarity matrix consisting of dissimilarity indexes, which quantitatively measure the extent that two vertices would like to be in the same sub-network, and then attempt to divide the nodes into clusters such that dissimilarity between objects within the same cluster is minimized, while that between objects from different clusters is maximised. The main discrimination of different clustering methods is that they use their own dissimilarity index (see [7] for a review of different dissimilarity indexes). Our decomposition algorithm is similar to that of [16] which is based on the bow-tie structure of metabolic networks, while a different dissimilarity index is used. The algorithm begins with decomposing the Core part into sub-networks that are still strongly connected, or most of the nodes are strongly connected. In the viewpoint of graph theory, nodes in the same strongly connected component are structurally equivalent in the sense that, (1) any pair of nodes within this component is mutually reachable from each other; (2) if an outside node can reach any node in this component, it can reach all of its nodes; (3) if a node of this component can reach any outside node, all nodes of this component can reach the same node.

In a directed graph, the distance $d_{ij}$ from node $i$ to node $j$ is defined as the number of arcs in the shortest directed path from $i$ to $j$ (in general, $d_{ij} \neq d_{ji}$.). The distance between all node pairs can be computed by Floyd algorithm [28]. For any vertex $i$, the set $\{d_{i1},...,d_{i,i-1},d_{i,i+1},...,d_{iN}\}$ measures how far all the other nodes are located from it, while the set $\{d_{1i},...,d_{i-1,i},d_{i+1,i},...,d_{Ni}\}$ measures the reverse distances, where N is the number of nodes in this network. If two nodes $i$ and $j$ would belong to the same cluster, the distance $d_{ik}$ and $d_{ki}$ ($k \neq i, j$) should be quite similar to the distance $d_{jk}$ and $d_{kj}$ respectively, while nodes $i$ and $j$ which satisfy the condition $d_{ij}$ equals $d_{ji}$ would fall into the same cluster. Therefore, the dissimilarity index $D(i,j)$ between vertex $i$ and $j$ can be defined as the corrected Euclidean-like dissimilarity [37],

$$D(i,j) = \sqrt{(d_{ij}-d_{ji})^2 + \sum_{\substack{k=1 \\ k \neq i,j}}^{N}[(d_{ki}-d_{kj})^2 + (d_{ik}-d_{jk})^2]} \qquad (1)$$

This dissimilarity is compatible with the structural equivalence of nodes within a strongly connected component, where the three terms of this equation quantitatively measure the three aspects of the equivalence, respectively.

Having obtained the dissimilarity indexes, Ward's clustering, a hierarchically agglomerative clustering method, is used to decompose the network [38]. This method starts with each node being its own cluster, then at each step, combines the two most similar clusters to form a new cluster, until all the nodes have been combined into one cluster. The algorithm produces a hierarchical clustering tree, or a dendrogram, for the network. Finally, the decomposition of the Core part can be expanded to the global network by using the "majority rule" proposed by Ma and Zeng [16]. The algorithm steps are presented as follows:
1. Remove all the linear branches of the GSC part and get the Core.
2. Decompose the Core of the GSC by Ward's clustering based on the dissimilarity index of equation (1) and get its hierarchical clustering tree.
3. Cut the hierarchical clustering tree into $m$ clusters so that the value of modularity metric is the largest [11].
4. Expand the clusters of the Core to the whole metabolic network by the "majority rule", i.e., the nodes that are directly connected to nodes in GSC are placed in the subset to which most of their neighbours in GSC belong; the other nodes are classified into corresponding subsets to which most of their neighbours belong.

It is worth to note that with the dissimilarity index of equation (1), nodes that belong to the same sub-tree own the highest degree of "structural equivalence", but are not necessarily connected to each other, see Figure S9(B) and S10(B) in Additional file 1 for an example. In part II of Additional file 1, we illustrate each step of this algorithm using a relatively small genome-based metabolic network of *Aeropyrum pernix* (*ape*).

In Additional file 3, we list metabolite abbreviations of *ape* network.

### Algorithm to generate random networks

A method similar to that of Maslov *et al.* [20, 21] was used to generate an ensemble of randomized networks. This algorithm randomly reshuffles the links of the original network, while preserving the in- and out-degree of each node, as well as the total number of directed and bi-directed arcs. It is presented as follows:

1. Partition all the arcs of the original network into directed arcs and bi-directed arcs.
2. Reshuffle the bi-directed arcs: Randomly select a pair of bi-directed arcs A↔B and C↔D. Rewire the two bi-directed arcs to get links A↔C and B↔D, if there are neither directed nor bi-directed arcs between the pair A-C and B-D respectively. Otherwise, abandon this pair and chose another pair of bi-directed arcs. The last restriction prevents the appearance of multiple arcs between the same pair of nodes. In addition, the network should always remain connected during the rewiring process.
3. Reshuffle the directed arcs: Randomly select a pair of directed arcs A→B and C→D. Rewire the two directed arcs to get links A→D and C→B. As in step 2, during the rewiring process, multiple links are prohibited, and the network should always remain connected.

Repeating step 2 and step 3 many times will generate a randomly connected counterpart of the original network.

### Method to compare a real metabolic network with randomized ones

Following the scheme of Maslov *et al.* [21], we apply Z-score to quantify the difference between a real metabolic network and its randomized counterparts:

$$Z = \frac{P - \overline{P}_r}{\Delta P_r},$$

where P is the graph metric in the real network, $\overline{P}_r$ and $\Delta P_r$ are the mean and standard deviation of the corresponding graph metric in the randomized ensemble.

## Authors' contributions

JZ conceived of the study, designed the analysis, implemented the analysis and prepared the manuscript. HY helped JZ to implement the analysis. JHL managed the project. ZWC and YXL helped JZ to design the analysis, provided guidance, coordinated and participated in the biological and theoretical analyses, and revised the manuscript. All authors read and approved the final manuscript.

## Acknowledgements

We thank Dr. R. Guimerà and Dr. L. A. N. Amaral for kindly providing us the software Modul-w to compute network modularity metric; Dr. H.W. Ma and Dr. A.P. Zeng for providing us with their metabolic networks' database; Dr. J. Doyle and Dr. Y. Z. Chen for critical reading and constructive comments; and the anonymous reviewers for their constructive comments. This work was supported in part by grants from Ministry of Science and Technology China(2003CB715900, 04BA711A21, 2004CB720103)，National Natural Science Foundation of China (30500107)，and Science and technology commission of Shanghai municipality（04DZ19850, 04DZ14005）.

# References

1. Stelling J, Klamt S, Bettenbrock K, Schuster S, Gilles ED: **Metabolic network structure determines key aspects of functionality and regulation.** *Nature* 2002, **420:**190-193.
2. Wagner A, Fell DA: **The small world inside large metabolic networks.** *Proc R Soc Lond B* 2001, **268:**1803-1810.
3. Jeong H, Tombor B, Albert R, Oltvai ZN, Barabasi AL: **The large-scale organization of metabolic networks.** *Nature* 2000, **407:**651-654.
4. Tanaka R: **Scale-Rich Metabolic Networks.** *Physical Review Letters* 2005, **94:**168101.
5. Arita M: **The metabolic world of Escherichia coli is not small.** *PNAS* 2004, **101:**1543-1547.
6. Arita M: **Scale-Freeness and Biological Networks.** *J Biochem (Tokyo)* 2005, **138:**1-4.
7. Zhao J, Yu H, Luo J, Cao Z, Li Y: **Complex networks theory for analyzing metabolic networks.** *Chinese Science Bulletin* 2006, **51:**1529-1537.
8. Erdos P, Renyi A: **On the evolution of random graphs.** *Publication of the Mathematical Institute of the Hungarian Academy of Science* 1960, **5:**17-61.
9. Hartwell LH, Hopfield JJ, Leibler S, Murray AW: **From molecular to modular cell biology.** *Nature* 1999, **402:**C47-C52.
10. Papin JA, Reed JL, Palsson BO: **Hierarchical thinking in network biology: the unbiased modularization of biochemical networks** *Trends in Biochemical Sciences* 2004, **29:**641-647.
11. Newman MEJ, Girvan M: **Finding and evaluating community structure in networks.** *Physical Review E* 2004, **69:**026113.
12. Ravasz E, Somera AL, Mongru DA, Oltvai ZN, Barabasi AL: **Hierarchical Organization of Modularity in Metabolic Networks.** *Science* 2002, **297:**1551-1555.
13. Gagneur J, Jackson DB, Casari G: **Hierarchical analysis of dependency in metabolic networks.** *Bioinformatics* 2003, **19:**1027-1034.
14. Guimera R, Amaral LAN: **Cartography of complex networks: modules and universal roles.** *Journal of Statistical Mechanics: Theory and Experiment* 2005**:**P02001
15. Holme P, Huss M, Jeong H: **Subnetwork hierarchies of biochemical pathways.** *Bioinformatics* 2003, **19:**532-538.
16. Ma H-W, Zhao X-M, Yuan Y-J, Zeng A-P: **Decomposition of metabolic network into functional modules based on the global connectivity structure of reaction graph.** *Bioinformatics* 2004, **20:**1870-1876.
17. Guimera R, Sales-Pardo M, Amaral LAN: **Modularity from fluctuations in random graphs and complex networks.** *Physical Review E* 2004, **70:**025101.
18. Guimera R, Nunes Amaral LA: **Functional cartography of complex metabolic networks.** *Nature* 2005, **433:**895-900.
19. Ma H-W, Zeng A-P: **The connectivity structure, giant strong component and centrality of metabolic networks.** *Bioinformatics* 2003, **19:**1423-1430.
20. Maslov S, Sneppen K: **Specificity and Stability in Topology of Protein Networks.** *Science* 2002, **296:**910-913.
21. Maslov S, Sneppen K, Zaliznyak A: **Detection of topological patterns in complex networks: correlation profile of the internet.** *Physica A: Statistical and Theoretical Physics* 2004, **333:**529-540.

22. Goto S, Nishioka T, Kanehisa M: **LIGAND: chemical database of enzyme reactions.** *Nucl Acids Res* 2000, **28:**380-382.
23. Goto S, Okuno Y, Hattori M, Nishioka T, Kanehisa M: **LIGAND: database of chemical compounds and reactions in biological pathways.** *Nucl Acids Res* 2002, **30:**402-404.
24. Csete M, Doyle J: **Bow ties, metabolism and disease.** *Trends in Biotechnology* 2004, **22:**446-450.
25. DANDEKAR T, SCHUSTER S, SNEL B, HUYNEN M, BORK P: **Pathway alignment: application to the comparative analysis of glycolytic enzymes.** *Biochemical Journal* 1999, **343:**115-124.
26. Huynenb MA, Dandekarb T, Bork P: **Variation and evolution of the citric-acid cycle: a genomic perspective** *Trends in Microbiology* 1999, **7:**281-291
27. Spirin V, Gelfand MS, Mironov AA, Mirny LA: **A metabolic network in the evolutionary context: Multiscale structure and modularity.** *PNAS* 2006, **103:**8774-8779.
28. Bondy JA, Murty USR: *Graph theory with applications.* London: Macmillan; 1976.
29. Watts DJ, Strogatz SH: **Collective dynamics of `small-world' networks.** *Nature* 1998, **393:**440-442.
30. Ma H, Zeng A-P: **Reconstruction of metabolic networks from genome data and analysis of their global structure for various organisms.** *Bioinformatics* 2003, **19:**270-277.
31. Duarte NC, Herrgard MJ, Palsson BO: **Reconstruction and Validation of Saccharomyces cerevisiae iND750, a Fully Compartmentalized Genome-Scale Metabolic Model.** *Genome Res* 2004, **14:**1298-1309.
32. Kitano H: **Biological robustness.** *Nature Reviews Genetics* 2004, **5:**826-837.
33. Tanaka R, Csete M, Doyle J: **Highly optimised global organisation of metabolic networks** *IEE Proceedings - Systems Biology* 2005, **152:**179-184.
34. Kitano H, Oda K: **Robustness trade-offs and host-microbial symbiosis in the immune system.** *Mol Syst Biol* 2006, **2:**E1-E10.
35. Kitano H, Oda K, Kimura T, Matsuoka Y, Csete M, Doyle J, Muramatsu M: **Metabolic Syndrome and Robustness Tradeoffs.** *Diabetes* 2004, **53:**S6-15.
36. Marhl M, Perc M, Schuster S: **Selective regulation of cellular processes via protein cascades acting as band-pass filters for time-limited oscillations.** *FEBS Letters* 2005, **579:**5461-5465.
37. Batagelj V, Mrvar A, Ferligoj A, Doreian P: **Generalized blockmodeling with Pajek.** *Metodoloski Zvezki* 2004, **1:**455-467

38. Ward J: **Hierarchical grouping to optimize an objective function.** *J Amer Statist Assoc* 1963, **58:**236-244.
39. Batagelj V, Mrvar A: **Pajek-program for large network analysis.** *Connections* 1998, **21:**47-57.

# Figure legends

**Figure 1 - The hierarchical clustering tree for the Core of the GSC for the *E.coli* network**

See Additional file 2 for metabolite abbreviations of the *E.coli* network.

**Figure 2 - Decomposition of the Core for the GSC of the *E.coli* metabolic network.**

This graph is drawn with the graph analysis software Pajek [39]. The nodes included in the biggest strongly connected component of each cluster are shown in red colour.

**Figure 3 - Decomposition of the *E.coli* metabolic network by expanding the clustering of the Core.**

Triangles correspond to the nodes of the Core. The four parts (GSC, S, P, IS) of bow-tie structure for the modules are shown in distinct colours.

**Figure 4 - Cartographic representation of the metabolic network for *E.coli*.**

Each circle represents a module and is coloured according to the KEGG pathway classification of the reactions belonging to it, while the arcs reflect the connection between clusters. The area of each colour in one circle is proportional to the number of reactions that belong to the corresponding metabolism. The width of an arc is proportional to the number of reactions between the two corresponding modules. For simplicity, bi-directed arcs are presented by grey edges.

**Figure 5 - Bio-reactions in the 3rd module and the connection to other modules.**

Each node represents a metabolite and is coloured according to the class of metabolism it participates in. This module contains the majority of metabolites from TCA cycle with glyoxylate bypass, in which the reactions are highlighted by red arcs. Nodes from other modules that link with module 3 are shown by triangles, with module serial number shown in the parentheses. The metabolite abbreviations are listed in Additional file 2.

**Figure 6 - Distribution of the 12 precursors in the 12 modules of the *E.coli* metabolic network.**

The three major pathways – Embden-Meyerhof-Parnas (EMP), tricarboxylic acid (TCA) and pentose phosphate pathway (PPP) for the generation of the 12 precursors are outlined.

**Figure 7 – Corresponding sub-tree and bow-tie structure of module 3.**
**(A) Sub-tree of module 3**           **(B) Bow-tie structure of module 3**
Each branch of the sub-tree corresponds to a red node in module 3, while the pink node titled "OASUC" also has parallelism in the sub-tree because it is included in the Core of *E.coli* network. These nodes were resulted from the decomposition of the Core. Then by the "majority role" the Core clusters were expanded to the whole network, the pink (other than "OASUC"), green, and blue nodes were assigned to cluster 3. The metabolite abbreviations are listed in Additional file 2.

**Figure 8 - The connections among the GSC parts of the twelve bow-tie like modules.**

The width of an arc is proportional to the number of links between the GSC parts of the two corresponding modules. For simplicity, bi-directed arcs are presented by grey edges.

**Figure 9 – Comparison of the Core of *E.coli* network with that of a randomized network.**

**(A) 12 clusters of the Core for *E.coli* network**

**(B) 12 clusters of the Core for a randomized network**

Both of the Cores are decomposed by our algorithm. Different clusters are shown in different colours. These two networks include 163 and 227 nodes respectively. The network in (A) and the decomposition result is just the same as that in Figure 2. The network in (B) is the Core of the 51$^{st}$ network in Table S4 of Additional file 1.

# Tables

**Table 1  Node distributions in the global structure of sub-networks obtained from the decomposition for *E.coli* network**

| Module | Total nodes | Nodes in GSC | Percent of GSC | Nodes in S | Percent of S | Nodes in P | Percent of P | Nodes in IS | Percent of IS | Bow-tie |
|---|---|---|---|---|---|---|---|---|---|---|
| 1 | 66 | 28 | 42% | 21 | 32% | 16 | 24% | 1 | 2% | Y |
| 2 | 60 | 23 | 38% | 1 | 2% | 27 | 45% | 9 | 15% | Y |
| 3 | 23 | 15 | 65% | 3 | 13% | 5 | 22% | 0 | 0 | Y |
| 4 | 44 | 15 | 34% | 14 | 32% | 4 | 9% | 11 | 25% | Y |
| 5 | 136 | 40 | 29% | 17 | 13% | 51 | 38% | 28 | 20% | Y |
| 6 | 21 | 13 | 62% | 6 | 29% | 0 | 0 | 2 | 9% | Y |
| 7 | 49 | 14 | 29% | 19 | 39% | 9 | 18% | 7 | 14% | Y |
| 8 | 19 | 8 | 42% | 8 | 42% | 3 | 16% | 0 | 0 | Y |
| 9 | 94 | 15 | 16% | 7 | 8% | 32 | 34% | 40 | 42% | Y |

| | | | | | | | | | |
|---|---|---|---|---|---|---|---|---|---|
| 10 | 28 | 7 | 25% | 6 | 21% | 5 | 18% | 10 | 36% | Y |
| 11 | 18 | 9 | 50% | 7 | 39% | 1 | 6% | 1 | 5% | Y |
| 12 | 17 | 10 | 59% | 1 | 6% | 6 | 35% | 0 | 0 | Y |
| Global network | 575 | 234 | 41% | 85 | 15% | 177 | 31% | 79 | 13% | Y |

**Table 2  Comparison of the *E.coli* metabolic network with sixty randomized networks.**

| | GSC | S | P | IS | Core | C | M |
|---|---|---|---|---|---|---|---|
| Mean of the sixty randomized networks | 287 | 90 | 126 | 71 | 205 | 0.0027 | 0.7601 |
| Standard deviation of the sixty randomized networks | 15.86 | 10.23 | 14.37 | 13.72 | 12.43 | 0.0019 | 0.0043 |
| E.coli network | 234 | 85 | 177 | 79 | 163 | 0.0646 | 0.8527 |
| Z-score | -3.40 | -0.52 | 3.53 | 0.61 | -3.37 | 31.91 | 21.79 |

C: Average clustering coefficient of the network

M: Modularity metric of the network obtained by simulated annealing algorithm

**Table 3  Node distributions in the sub-networks of the cell compartment reactions for *S. cerevisiae***

| Sub-network | Total nodes | Nodes in GSC | Percent of GSC | Nodes in S | Percent of S | Nodes in P | Percent of P | Nodes in IS | Percent of IS | Bow-tie |
|---|---|---|---|---|---|---|---|---|---|---|
| [c] | 427 | 206 | 48.24% | 33 | 7.73% | 154 | 36.07% | 34 | 7.96% | Y |
| [m] | 72 | 35 | 48.61% | 9 | 12.50% | 26 | 36.11% | 2 | 2.78% | Y |
| [x] | 48 | / | / | / | / | / | / | / | / | N |
| Global network | 556 | 269 | 48.38% | 39 | 7.01% | 229 | 41.19% | 19 | 3.42% | Y |

Compartment Abbreviations

[c]: cytosol; [m]: mitochondrion;   [x] : peroxisome

# Additional files

**Additional file 1 –supplementary 1.pdf**
Supplementary material for this paper
**Additional file 2 –supplementary 2_eco_matabolite abbreviation.xls**
List of abbreviations for the metabolites in *E.coli* network
**Additional file 3 –supplementary 3_ape_matabolite abbreviation.xls**
List of abbreviations for the metabolites in *ape* network

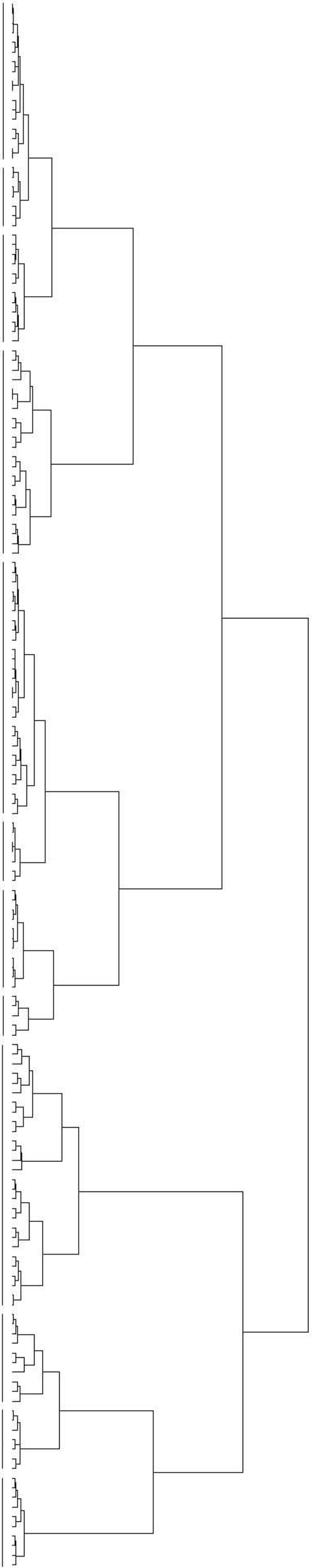

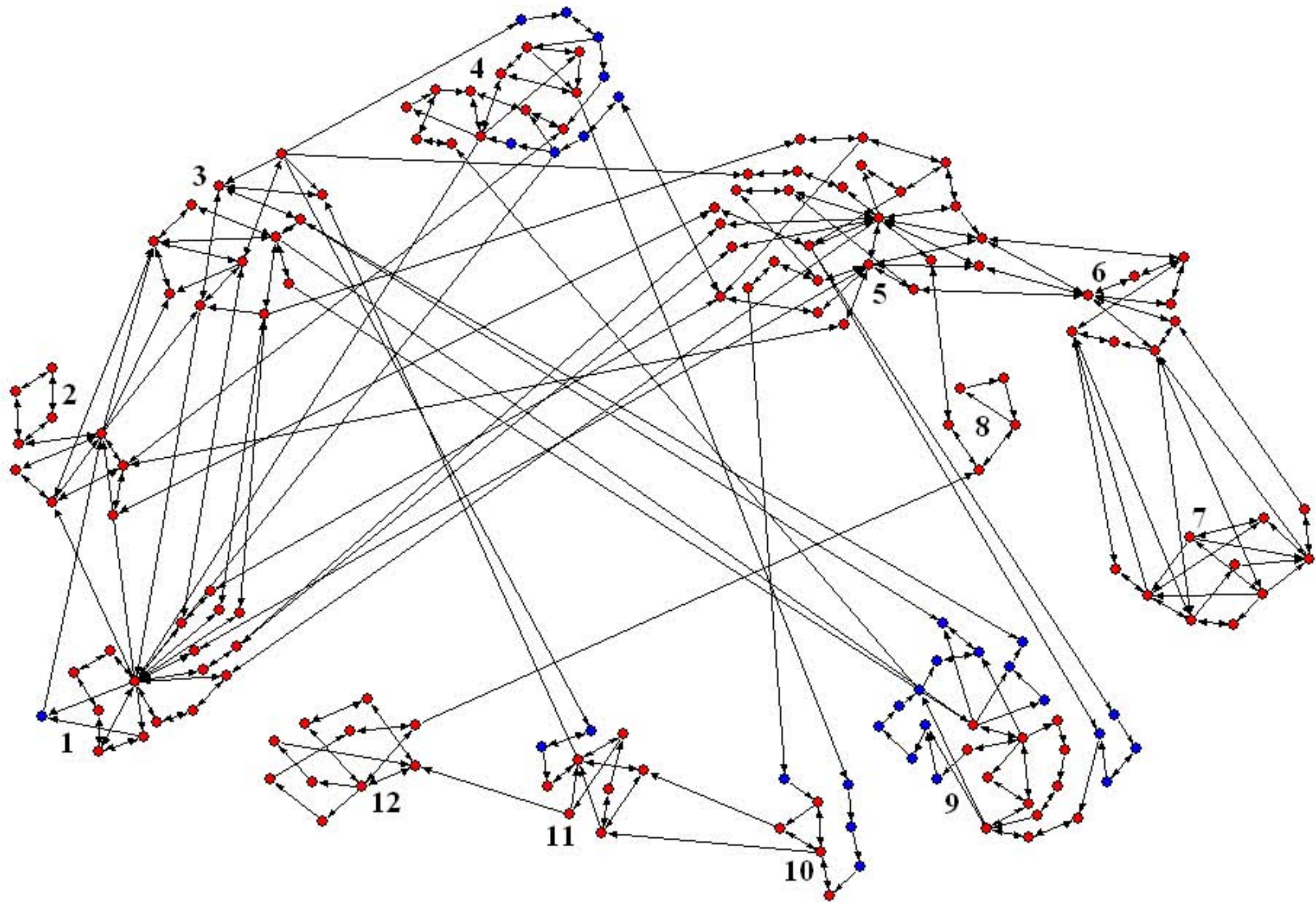

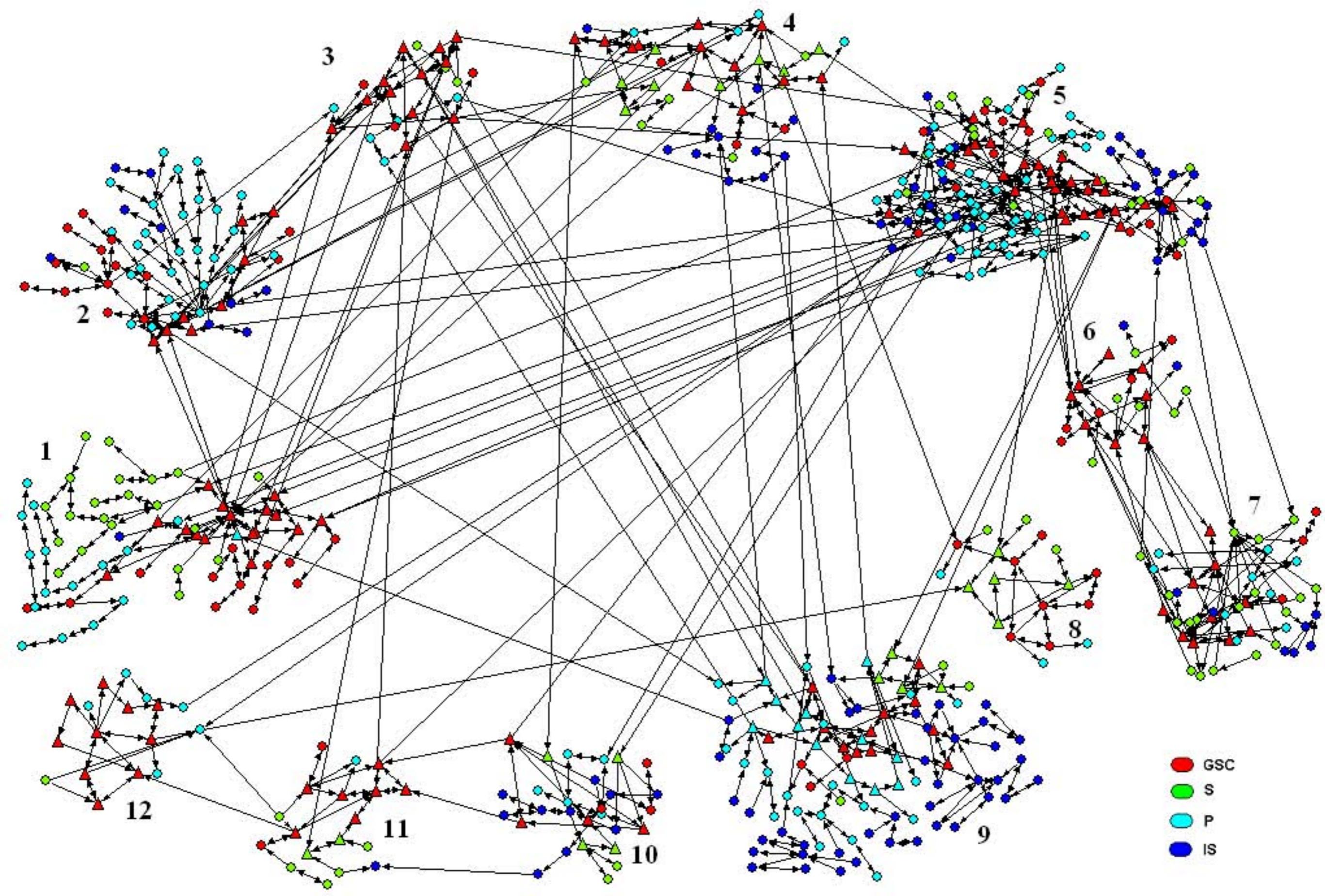

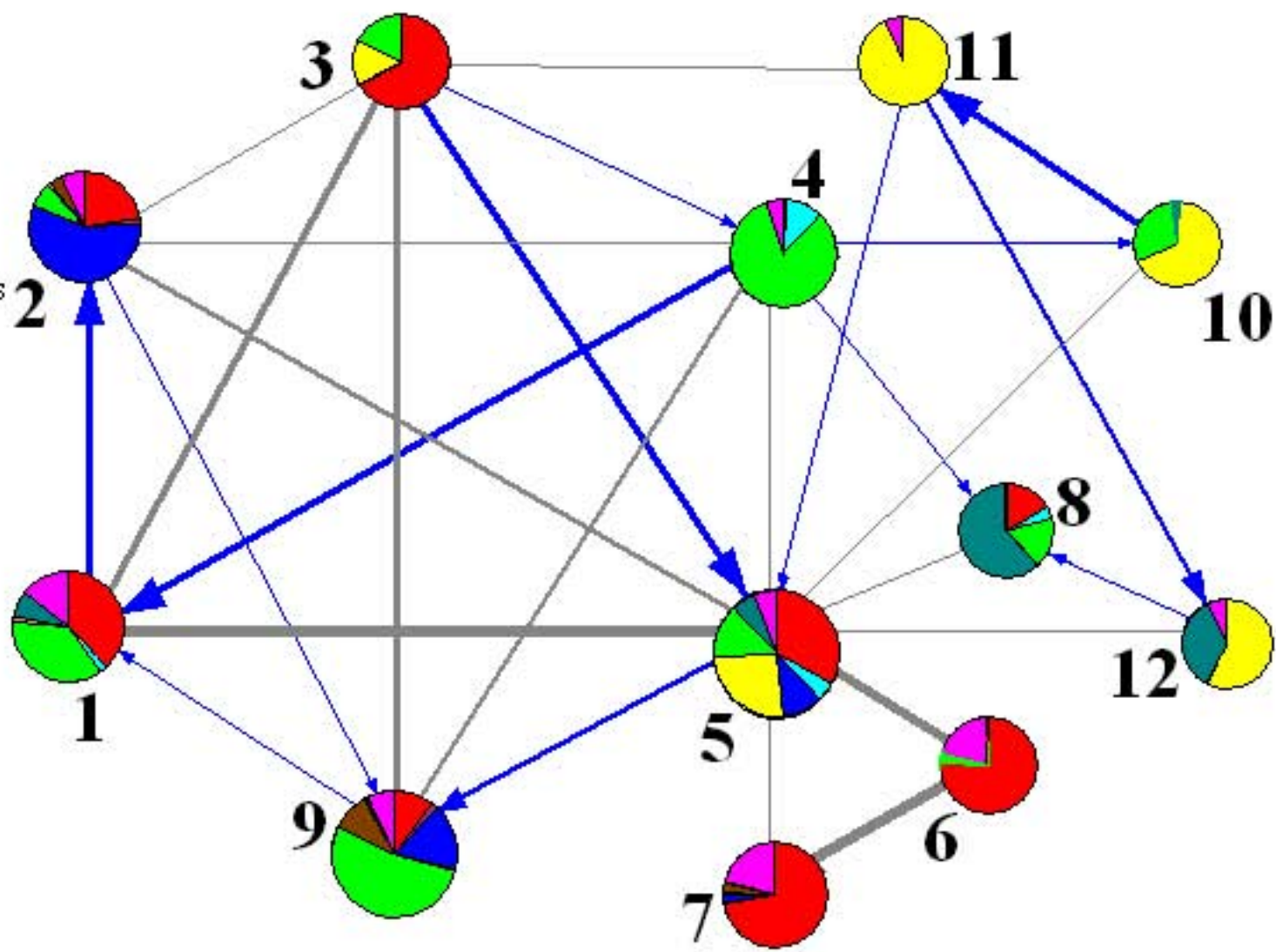

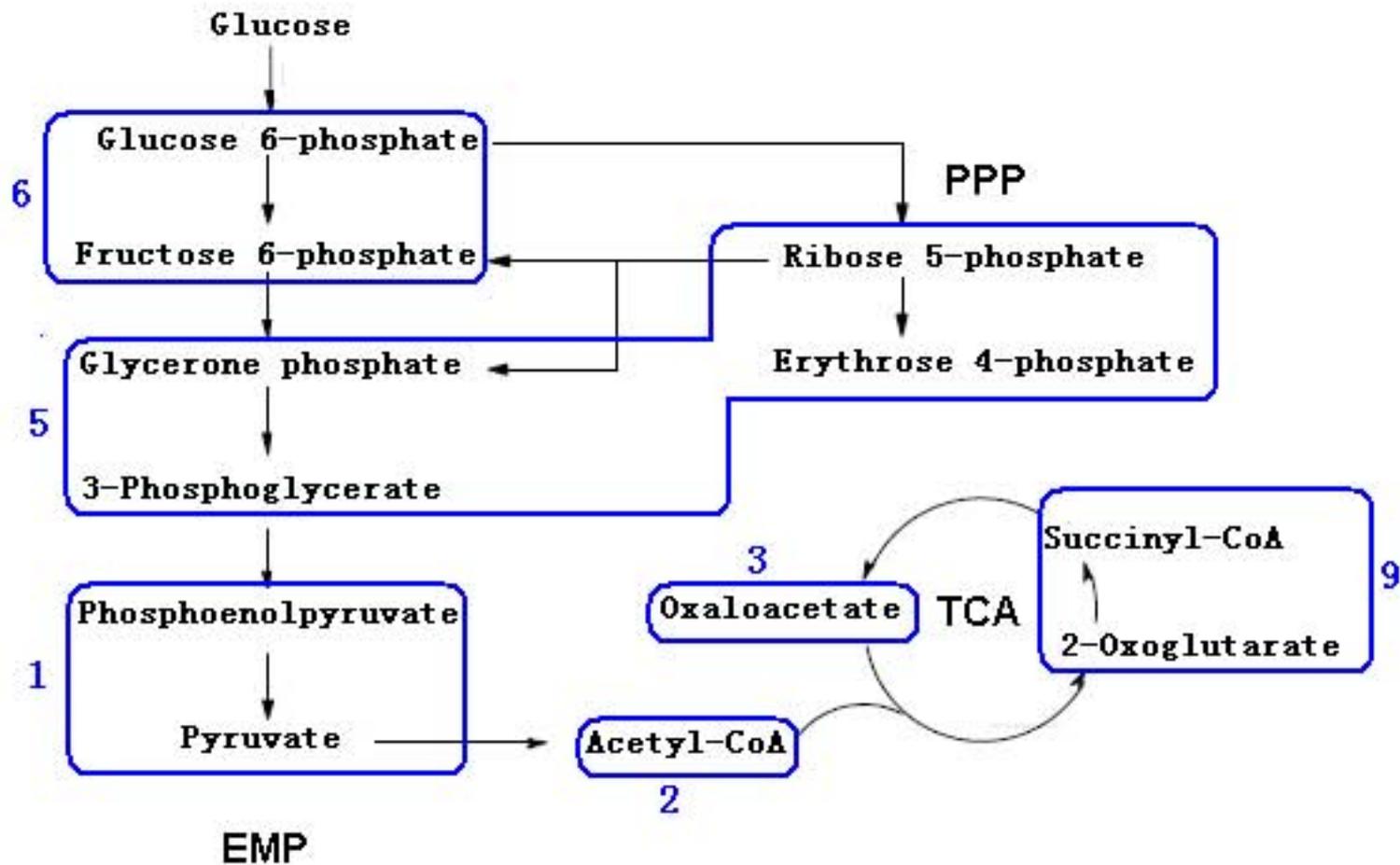

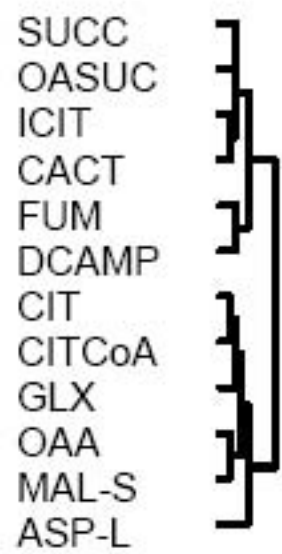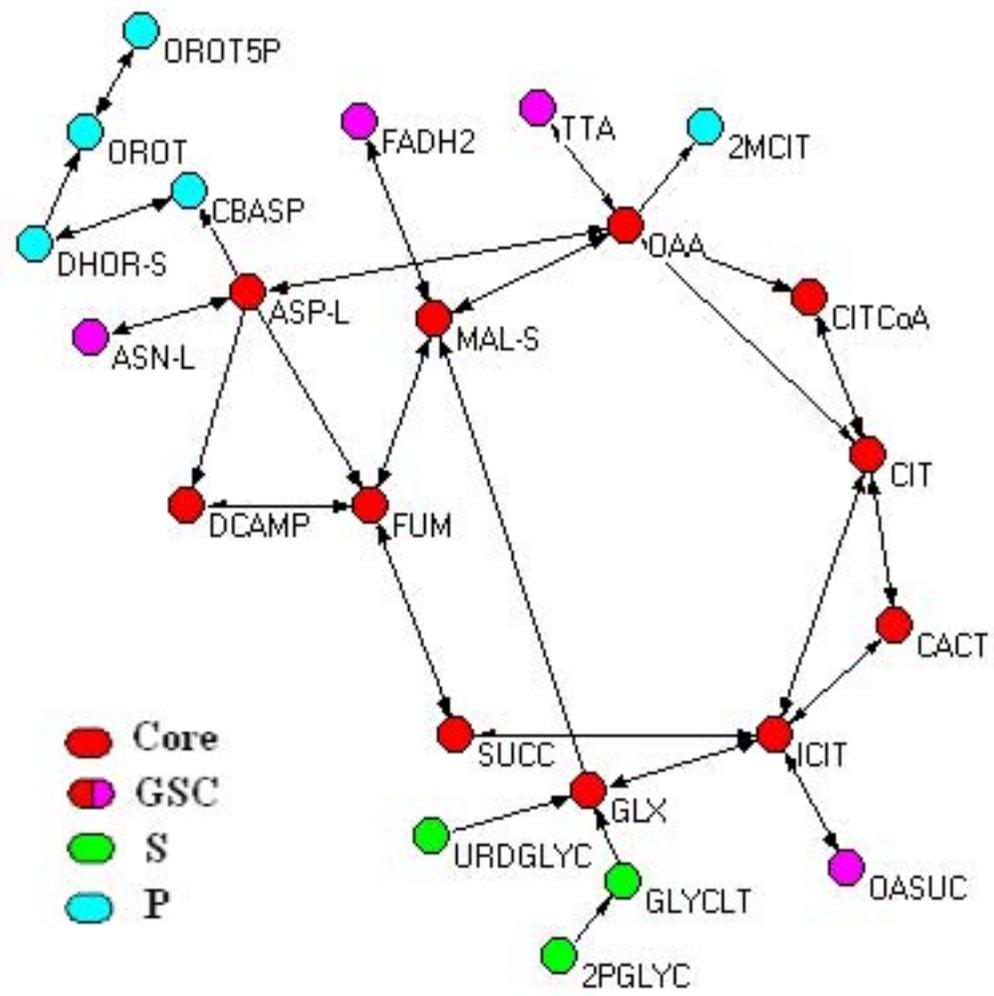

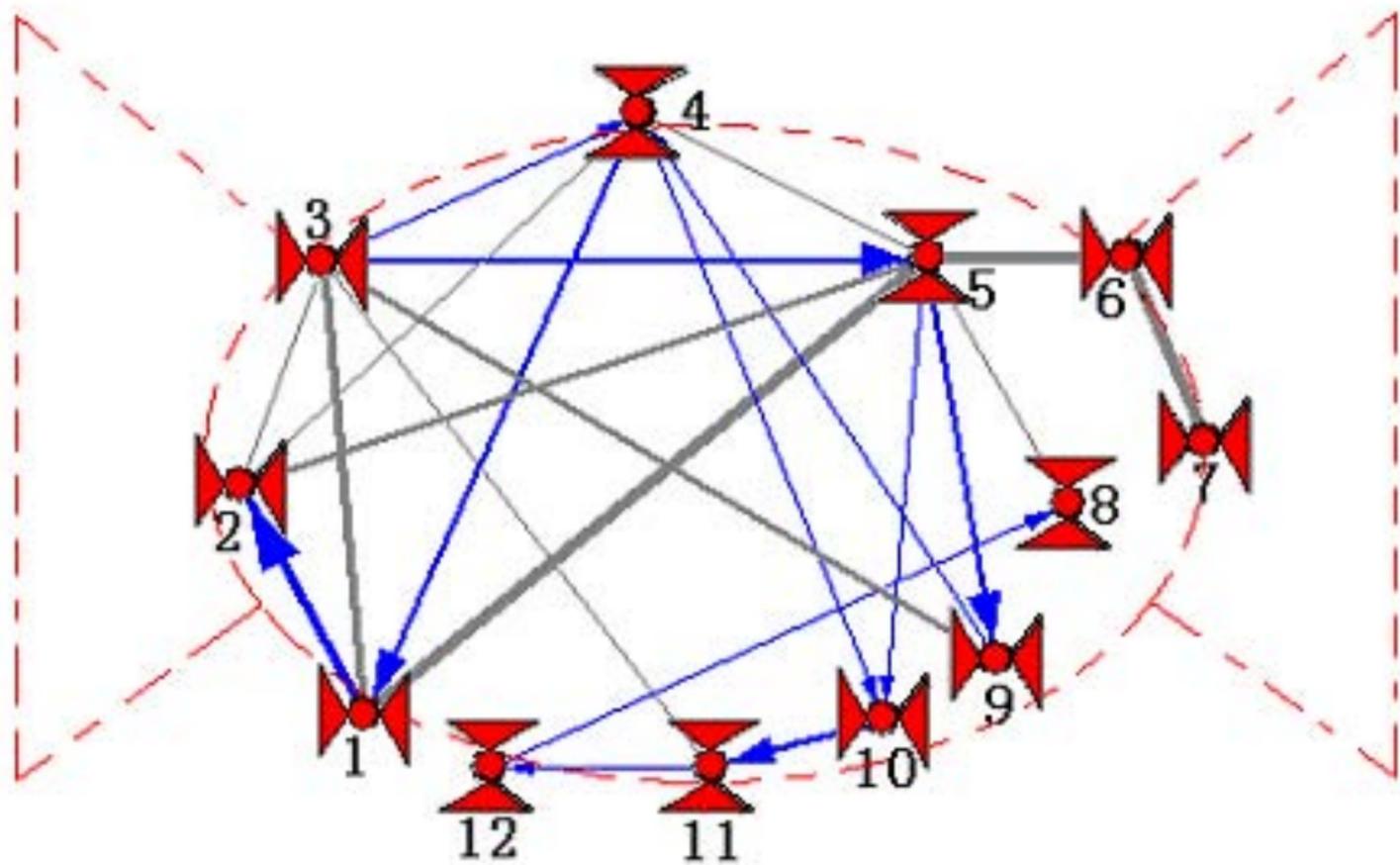

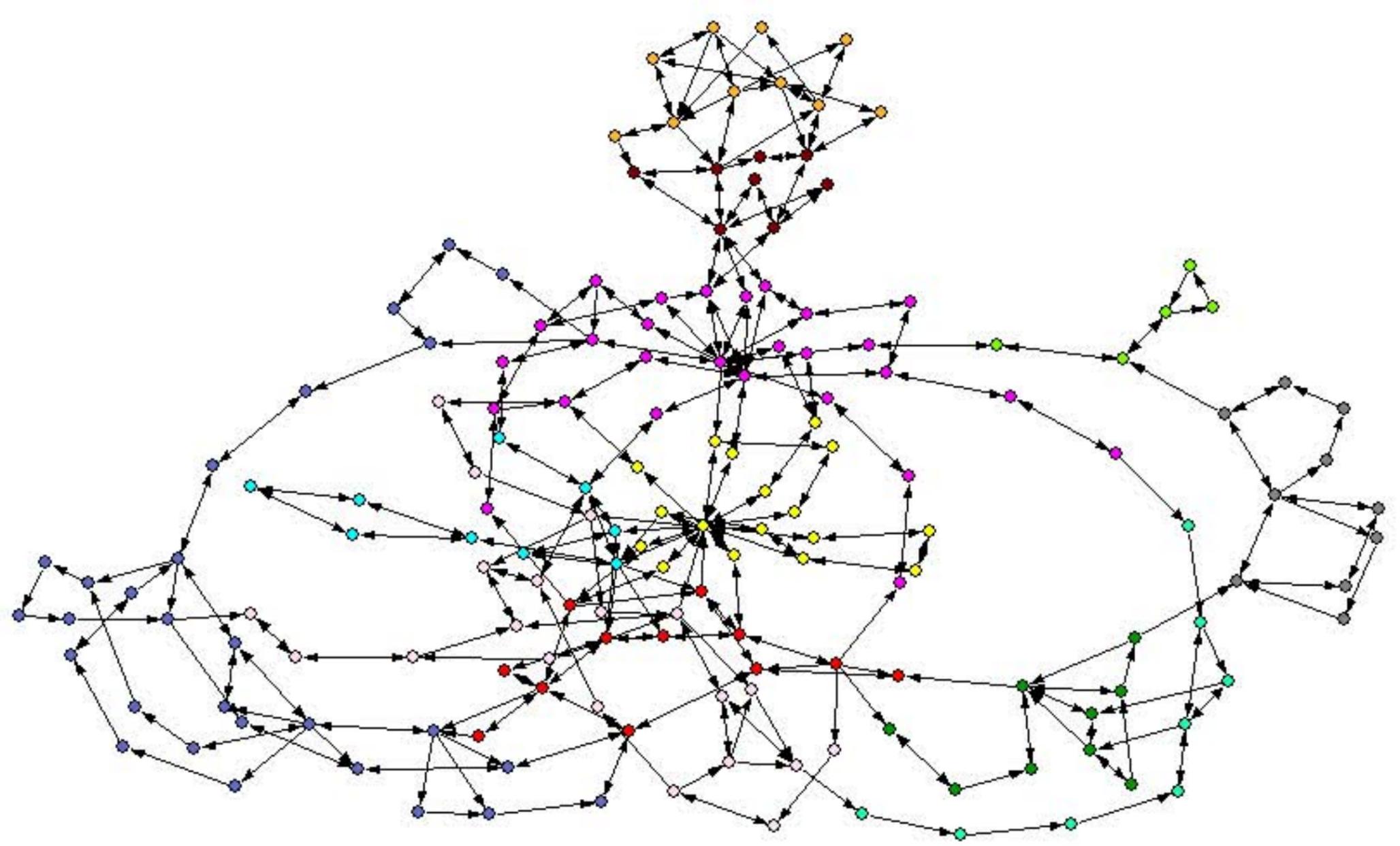

**(A)**

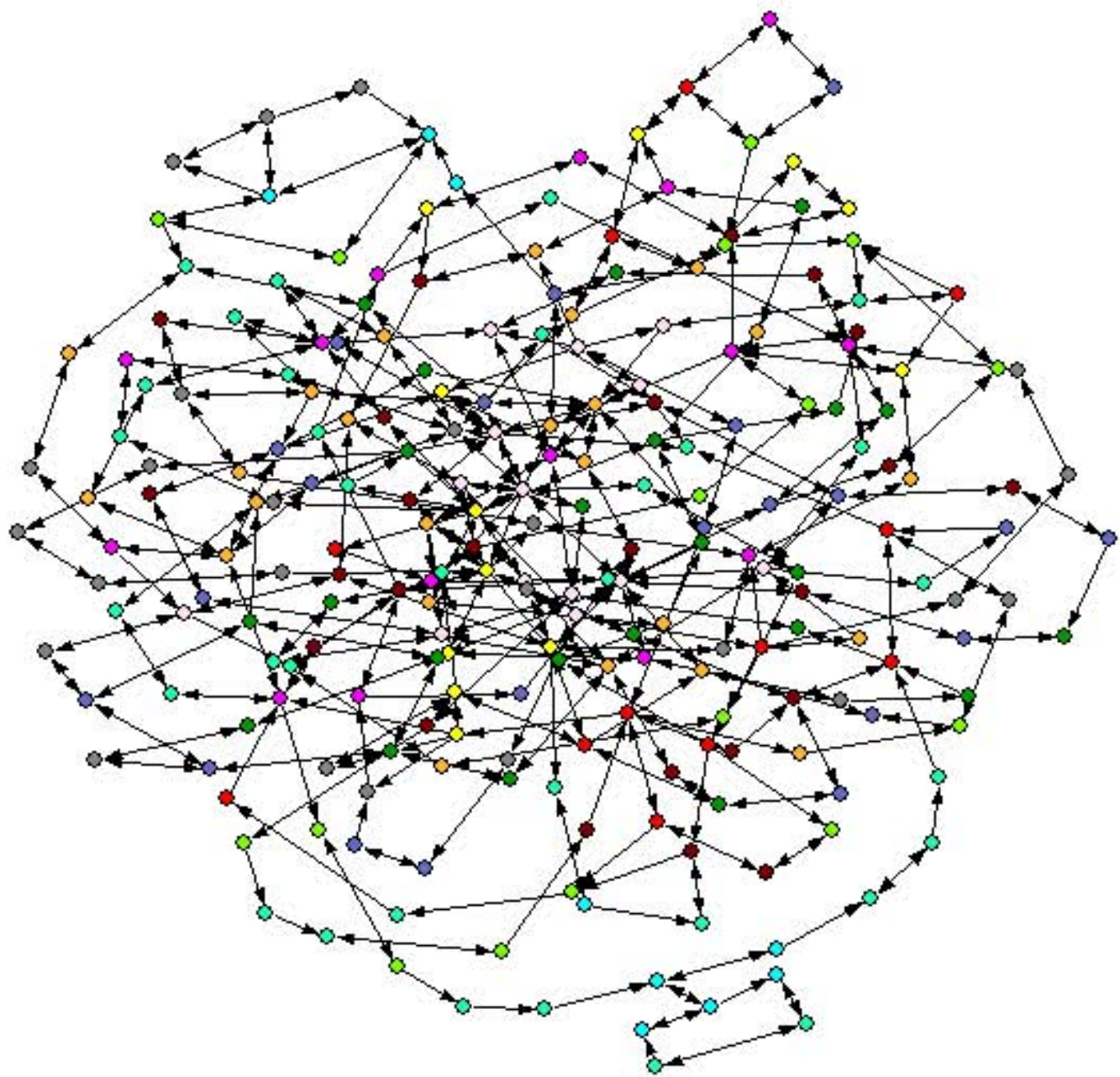

**(B)**